\documentclass[twocolumn,amsmath,amssymb]{revtex4-2}

\usepackage{graphicx}
\usepackage[utf8]{inputenc}
\usepackage{amsmath,amssymb}
\usepackage{epstopdf}
\usepackage{hyperref}
\usepackage{lineno}
\usepackage{color}

\newcommand{\Fref}[1]{Fig.~\ref{#1}}
\newcommand{\Eqref}[1]{Eq.~(\ref{#1})}
\newcommand{\be}{\begin{equation}}
\newcommand{\ee}{\end{equation}}
\newcommand{\bal}{\begin{align}}
\newcommand{\eal}{\end{align}}
\newcommand{\bear}{\begin{eqnarray}}
\newcommand{\eear}{\end{eqnarray}}
\newcommand{\nn}{\nonumber}

\newcommand{\md}{\mathrm{d}}
\newcommand{\e}{\mathrm{e}}

\newcommand{\kb}{k_\mathrm{_B}}
\newcommand{\alf}{{Alfv\'en~}}

\newcommand{\AcousticHeating}{Kalkofen:07,Tu:13,ChromoView:19,Pelekhata:23,Molnar:23,Judge:24,Vytenis:25,Udnaes:25}

\begin{document}

\title{On the Theory of Absorption of Sound Waves via the\\ Bulk Viscosity in the Partially Ionized Solar Chromosphere}

\author{Albert~M.~Varonov}
\email[E-mail: ]{varonov@issp.bas.bg}
\affiliation{Georgi Nadjakov Institute of Solid State Physics, Bulgarian Academy of Sciences,\\
72 Tzarigradsko Chaussee Blvd., BG-1784 Sofia, Bulgaria}

\author{Todor~M.~Mishonov}
\email[E-mail: ]{mishonov@gmail.com}
\affiliation{Georgi Nadjakov Institute of Solid State Physics, Bulgarian Academy of Sciences,\\
72 Tzarigradsko Chaussee Blvd., BG-1784 Sofia, Bulgaria}

\email{mishonov@gmail.com}

\date{1 January 2026}

\begin{abstract}
Bulk viscosity and thermodynamic variables of a hydrogen-helium cocktail: internal energy, enthalpy, pressure, their derivatives, heat capacities per constant density and pressure are obtained using temperature and density height profiles of the solar atmosphere~[Avrett \& Loeser, ApJS Vol.~175, 229 (2008)].
The qualitative evaluation for the necessary sound wave energy flux 
to heat the solar chromosphere is determined to be 320~kW/m$^2$.
It is concluded that the bulk viscosity creates the dominating mechanism 
of acoustic waves damping and it is not necessary to introduce artificial 
viscosity or to conclude that shear viscosity is not sufficient for chromosphere heating; the volume viscosity induced wave absorption is sufficient.
\end{abstract}

\maketitle

\section{Introduction}

Little or almost no attention has been given to the influence of the bulk or volume viscosity in the physical processes in the solar atmosphere.
While this is reasonable in the outer part, the solar corona, where the enormous temperatures lead to full hydrogen ionization and hence zero bulk viscosity, for the inner part up to the transition region, the solar chromosphere, the partially ionized low temperature plasma has a bulk viscosity orders of magnitude larger than the shear one.
Therefore the absorption of sound (acoustic) waves is much larger than the same of the magneto-hydrodynamic transversal waves (slow magneto-sonic and \alf waves) and therefore much more energy from sound waves is deposited in the solar chromospheric partially ionized plasma. 

Sound or acoustic wave propagation, absorption and consequent heating in the solar chromosphere is a widely studied topic~\cite{\AcousticHeating}.
However, the main mechanism of sound wave absorption in the solar chromosphere, namely via the bulk or volume viscosity $\zeta$ has received little to no attention so far.
For instance, the numerical simulations of the solar corona in Ref.~\cite{Einaudi:21} has bulk viscosity included but its value is set to 2/3 of that of the shear one.
Another magneto-hydrodynamic code includes and considers the bulk as an artificial one~\cite{Roth:22}.
For mono-atomic gases $\zeta=0$~\cite[Chap.~1, Sec.~8 Viscosity in the gas, Eq.~(8.17)]{LL10} but for the partially ionized solar chromospheric plasma where ionization-recombination processes take place, it is the dominant mechanism for sound wave absorption~\cite[Fig.~1]{PoP:20}.
An \textit{ab-initio} study of the bulk viscosity due to ionization-recombination processes in local thermodynamic equilibrium (LTE) approximation in pure hydrogen plasma was completed not so long ago~\cite{PhysA} and recently it has been complemented with a general one for cold plasmas with temperatures much smaller than the ionization potentials of the constituent chemical elements~\cite{PoP:26}.
And now, the all the necessary ingredients to theoretically study the propagation and absorption of sound waves in the quiet solar chromosphere are present and this one of the first steps in doing so, as the quiet solar atmosphere is in LTE in an excellent approximation~\cite{ApJ:24}.

Contemporary simulation results from observed acoustic fluxes point that the latter are likely insufficient to heat the quiet solar chromosphere~\cite{Abbasvand:20a,Abbasvand:20b,Molnar:23}.
It is generally noted that another mechanism is necessary to heat the solar chromosphere and here it is shown that no new mechanism is needed, simply sound absorption through the bulk viscosity has to be accounted for.
And precisely in this region the plasma temperatures are lower and therefore both shear viscosity $\eta \propto T^{5/2}$~\cite[Eq.~(42.10)]{LL10} and thermal conductivity $\varkappa \propto T^{5/2}$~\cite[Eq.~(43.9)]{LL10} have correspondingly small values, the proton and electron temperatures are approximately equal.

\subsection{Brief Historic Quanta}

We will repeat in short the main idea of the present study.
Hannes Alfv\'en was one of first who consider that waves (and magneto-hydrodynamic MHD waves)
can heat the solar atmosphere~\cite{Alfven:42,Alfven:47}; that is why every list of references in incomplete.
Solar plasma is a weakly magnetized fluid and what can propagate in a fluid:
waves.
How can waves heat the fluid: by a kinetic coefficient describing entropy 
production in a local thermodynamic equilibrium in the first approximation.
How many dissipative coefficients can a fluid have:
actually not so many: heat $\varkappa$ and Ohmic $\sigma_{_\Omega}$ conductivity, shear $\eta$ and volume $\zeta$ viscosity.
Many authors analyzing acoustic heating of chromosphere conclude that
$\eta$ is not sufficient and even introduce in some cases some \textit{ad hoc} artificial viscosity.
The purpose of the present study is to illustrate that all those studies are in a correct
track, almost correct and only the bulk viscosity has to be incorporated in their considerations.
For the hydrogen atom it is all simple: the analytical energy spectrum and near threshold ionization cross-section by electron impact.
Incorporation of a cross-section~\cite{Wannier:53} and \cite[Chap.~18, Sec.~147 Behaviour of cross-sections near the reaction threshold]{LL3} for calculation of a kinetic coefficient in a gas
is a routine task of the statistical physics and for pure hydrogen plasma and even
for a cold plasma cocktail it is already a solved problem which we use
as a basis of our theory for chromosperic heating by acoustic waves.

\section{Recalling hydrodynamics of sound waves}

\subsection{Energy and momentum of a sound wave}

Let in the beginning analyze the propagation of a longitudinal sound wave 
with $x$-component of the velocity
\be
v(t,x)=v_0\cos(k^\prime x-\omega t),
\quad 
\omega=ck^\prime,
\quad
\langle v^2 \rangle=\frac12v_0^2,
\ee
where $c$ is sound velocity.
The averaged energy density \cite[Eq.~(65.3)]{LL6} 
\be
E=\rho \langle v^2 \rangle,
\ee
and energy flux density in $x$-direction $q=q_x$
\be
q=cE=\Pi.
\ee
This energy flux density is equal to the $x$-component of the momentum flux in $x$-direction 
$\Pi=\Pi_{xx}$.

In short wavelength (WKB) approximation, i.e. for weak 
$k^{\prime\prime}\ll k^{\prime}$ dumping rate
$v_0\propto\e^{-k^{\prime\prime}x}$ in approximately homogeneous fluid
$\rho\approx\mathrm{const}$
wave damping creates force in $x$-direction with volume density 
\be
f=-\md_x \Pi(x)=2k^{\prime\prime}q=Q_\zeta,
\qquad 
\md_x= \md/\md x,
\label{force_density_and-heating rate}
\ee
see \cite[Eq.~(7.1)]{LL6}
Simultaneously this force density is equal to the volume density of 
heating power by wave damping $Q_\zeta$.
The space damping rate \cite[Eq.~(79.6)]{LL6}
\be
k^{\prime\prime}(\omega)
=\frac{\omega^2}{2\rho c^3}\left[
\left(\frac43 \eta+\zeta^\prime(\omega)\right)
+\left(\frac{1}{\mathcal C_v}-\frac{1}{\mathcal C_p}\right)\varkappa
\right]
\label{complete_damping}
\ee
depends on shear $\eta$ and frequency dependent bulk viscosity 
$\zeta^\prime(\omega)$,
thermal conductivity and heat capacities per unit mass
at constant volume $\mathcal C_v$ and pressure $\mathcal C_p$.

We consider sound waves with propagating in vertical $x$-direction 
and for the static gradient of pressure we have the the hydrostatic equation
\be
\md_x p=f-g\rho,
\ee
where $g$ is the acceleration.

Simultaneously energy conservation gives
\be
\md_x\left(\frac12U^2+w+gx\right)=\frac{Q_\zeta-Q_r}{j},
\label{enthalpy_derivatives}
\ee
where $Q_r$ is the volume density of the radiative cooling,
$U$ is the very small in the chromosphere velocity of solar wind
and 
\be
j=\rho U=\mathrm{const}
\ee
is the mass flux of the solar wind.
For negligible dissipation the expression in parenthesis corresponds to the 
Bernoulli theorem \cite[Eq.~(5.4)]{LL6}.
The enthalpy $w$ per unit mass dominates in the parenthesis and
potential and kinetic energy per unit mass are negligible for the 
conditions in solar chromosphere.

In such a way 
supposing that energy flux related to thermal conductivity
$q_\varkappa=-\varkappa \md_xT$  is negligible,
we obtained an approximate system of 
equations
\begin{align}
\begin{pmatrix}
\md_x w\\
\md_x p
\end{pmatrix}
=\begin{pmatrix}
g(x)\\
f(x)
\end{pmatrix}
\equiv\begin{pmatrix}
(2k^{\prime\prime}q-Q_r)/j\\
2k^{\prime\prime}q-g\rho
\end{pmatrix}.
\label{system}
\end{align}
In the next subsection we will recall results for the thermodynamics 
and kinetics of the cold two component plasma.

\subsection{Thermodynamics and kinetic coefficients of cold H-He plasma}
Solar chromosphere we approximate as cold cocktail of H-He plasma.
Let $n_\rho$ is the volume density of hydrogen atoms.
For every hydrogen atom we have approximately 
$\overline a_\mathrm{He}\approx 0.1$ helium atoms.
And the mass density can be represented as
\be
\rho=M^*n_\rho,\qquad M^*=M+\overline a_\mathrm{He}M_\mathrm{He},
\ee
where $M$ is the proton mass and $M_\mathrm{He}\approx 4M$ 
is the mass of an alpha particle.

The temperature of $T$ order of $\frac12$eV is much lower than hydrogen
ionization energy 1 Rydberg, $I=\mathrm{R}=13.6$~eV.
At these conditions hydrogen is partially ionized with degree
of ionization
\be
\alpha=n_p/n_\rho,
\qquad n_\rho=n_p+n_0,
\ee
where $n_p$ is the volume density of protons and 
$n_0$ is the density of neutral hydrogen atoms.
The ionization of helium is negligible $\alpha_\mathrm{He}\approx 0$
and practically all electrons are created by hydrogen ionization
$n_e\approx n_p$.
The total number of particle per unit volume is
\begin{align}
&
n_\mathrm{tot}=(n_0+n_p)+n_e+n_\mathrm{He}
= \mathcal{N}_\mathrm{tot}n_\rho,\\
&
\mathcal{N}_\mathrm{tot}=1+\alpha+a_\mathrm{He}
\qquad
n_e=n_p=n_\rho\alpha.
\end{align}
And for the pressure in the approximation of ideal gas we have
\be
p= n_\mathrm{tot}T= \mathcal{N}_\mathrm{tot}n_\rho T
=(1+\alpha+a_\mathrm{He})n_\rho T.
\label{pressure_init}
\ee

For the hydrogen ionization the solution of Saha equation gives
\begin{align}
&
\alpha(\rho,T)
= f(\nu)\equiv\dfrac{2}{1+\sqrt{1+4\nu}}, \\
&
\nu\equiv\dfrac{n_\rho}{n_\mathrm{_S}} = \frac{1-\alpha}{\alpha^2}, 
\qquad
\frac1{\sqrt{1+4\nu}}=\frac{\alpha}{2-\alpha}
\\
&
n_\mathrm{_S}(T) \equiv n_q \e^{-\iota},
\quad
n_q(T) = \left(\frac{mT}{2\pi\hbar^2}\right)^{\!\!3/2},
\quad
\iota\equiv\frac{I}{T},
\end{align}
where $m$ is the electron mass and $n_\rho=\rho/M^*$,

For the enthalpy $w$ 
and internal energy $\varepsilon$
per unit mass we have \cite[Eqs.~(11-12), Eq.~(6)]{ApJ:21,ApJ:24}
\begin{align}
w(\rho,T)&=\frac{1}{M^*}\left(c_p\mathcal{N}_\mathrm{tot}T+I\alpha\right),
\qquad
c_p=\frac52,\\
\varepsilon(\rho,T)&=\frac{1}{M^*}\left(c_v\mathcal{N}_\mathrm{tot}T+I\alpha\right),
\qquad
c_v=\frac32.
\end{align}
For the derivatives we have
\begin{align}
f^\prime(\nu) & = \md_\nu f(\nu)
=-\frac4{\left(1+\sqrt{1+4\nu}\right)^{\!2}\sqrt{1+4\nu}}
\nn \\
& =\frac{-f^2(\nu)}{\sqrt{1+4\nu}}
=\frac{-f^3}{2-f}=-\frac{\alpha^3}{2-\alpha}=\frac{\md \alpha}{\md \nu},
\\
\nu f^\prime(\nu) & =-\frac{(1-\alpha)\alpha}{2-\alpha}=-\mathcal{D},
\qquad \mathcal{D}\equiv \frac{(1-\alpha)\alpha}{2-\alpha},
\label{cal_D}
\\
\left(\frac{\partial \alpha}{\partial T}\right)_{\!\!\rho} & =-\frac{(c_v+\iota)}{T}\nu f^\prime(\nu)
=\frac{(c_v+\iota)}{T}\frac{(1-\alpha)\alpha}{2-\alpha}\nn\\
&=\frac{(c_v+\iota)}{T}\mathcal{D},
\\
\left(\frac{\partial \alpha}{\partial \rho}\right)_{\!T} & = \nu f^\prime(\nu)/(M^*n_\mathrm{_S})
=-\frac{1}{M^*n_\mathrm{_S}} \frac{(1-\alpha)\alpha}{2-\alpha}\nn\\
& =-\frac{1}{M^*n_\mathrm{_S}} \mathcal{D}.
\end{align}
In such a way we obtain for the heat capacity the explicit expressions
\begin{align}
\mathcal{C}_v & \equiv\left(\frac{\partial \varepsilon}{\partial T}\right)_{\!\!\rho}
=\frac{1}{M^*}\left[c_v\mathcal{N}_\mathrm{tot}+ (c_v +\iota)^2  \mathcal{D}\right]
\\
& = \frac{1}{M^*}
\left[c_v(1+\alpha+\overline{a}_\mathrm{He})
+ (c_v +\iota)^2\frac{(1-\alpha)\alpha}{2-\alpha}\right].\nn
\end{align}

For derivatives of the pressure \Eqref{pressure_init} we have
\begin{align}
\left(\frac{\partial p}{\partial T}\right)_{\!\!\rho}
&=\left[\mathcal{N}_\mathrm{tot}+(c_v+\iota)\mathcal{D}\right]n_\rho \\
&=\left[ (1+\alpha+\overline{a}_\mathrm{He})+(c_v+\iota)
\frac{(1-\alpha)\alpha}{2-\alpha}\right]n_\rho,\nn \\
\left(\frac{\partial p}{\partial \rho}\right)_{\!\!T}
&=\frac{T}{M^*}\left[\mathcal{N}_\mathrm{tot}-\mathcal{D}\right] \\
&=\!\frac{T}{M^*}
\left[ (1+\alpha+\overline{a}_\mathrm{He})-
\frac{(1-\alpha)\alpha}{2-\alpha}\right]. \nn
\end{align}

Heat capacity per constant volume and unit mass $\mathcal{C}_v$
can be obtained by the general formula \cite[Eq.~(16.10)]{LL5}
in which volume is for unit mass $\mathcal{V}=1/\rho$
\begin{align}
\Delta\mathcal{C} & \equiv
\mathcal{C}_p-\mathcal{C}_v
=-T\frac{(\partial p/\partial T)_\mathcal{V}^2}
{(\partial p/\partial \mathcal{V})_T}
=\frac{T}{\rho^2}
\frac{(\partial p/\partial T)_\rho^2}
{(\partial p/\partial \rho)_T}\nn\\
&=\frac{\left[\mathcal{N}_\mathrm{tot}+(c_v+\iota)\mathcal{D}\right]^2}
{M^*\left[\mathcal{N}_\mathrm{tot}-\mathcal{D}\right]}\\
&=\dfrac{\left[ (1+\alpha+\overline{a}_\mathrm{He})+(c_v+\iota)
\dfrac{(1-\alpha)\alpha}{2-\alpha}\right]^2}
{M^*\left[ (1+\alpha+\overline{a}_\mathrm{He})-
\dfrac{(1-\alpha)\alpha}{2-\alpha}\right]},
\end{align}
and
\begin{align}
\mathcal{C}_p=&
\frac{\left[c_v\mathcal{N}_\mathrm{tot}+ (c_v +\iota)^2  \mathcal{D}\right]}{M^*}
+\frac{\left[\mathcal{N}_\mathrm{tot}+(c_v+\iota)\mathcal{D}\right]^2}
{M^*\left[\mathcal{N}_\mathrm{tot}-\mathcal{D}\right]}\\
=&\frac{\left[c_p\mathcal{N}_\mathrm{tot}+(c_vc_p+2c_p\iota+\iota^2)\mathcal{D}\right]\mathcal{N}_\mathrm{tot}}{(\mathcal{N}_\mathrm{tot}-\mathcal{D})M^*}
\\
= &\frac{1}{M^*}
\left[c_v(1+\alpha+\overline{a}_\mathrm{He})
+ (c_v +\iota)^2\frac{(1-\alpha)\alpha}{2-\alpha}\right]\nn\\
&+
\dfrac{\left[ (1+\alpha+\overline{a}_\mathrm{He})+(c_v+\iota)
\dfrac{(1-\alpha)\alpha}{2-\alpha}\right]^2}
{M^*\left[ (1+\alpha+\overline{a}_\mathrm{He})-
\dfrac{(1-\alpha)\alpha}{2-\alpha}\right]}.
\end{align}
For the coefficient of heat conductivity damping in \Eqref{complete_damping}
we have explicit and computable expression
\be
\frac{1}{\mathcal C_v}-\frac{1}{\mathcal C_p}
=\frac{\Delta\mathcal{C}}
{\mathcal C_v\mathcal C_p}.
\ee
These results generalize the previously obtained heat capacities~\cite{ApJ:21}
for pure hydrogen.  

Analogously for the derivatives of enthalpy
\begin{align}
\left(\frac{\partial w}{\partial T}\right)_{\!\!\rho}
& =\frac1{M^*}
\left[c_p\mathcal{N}_\mathrm{tot}+(c_v+\iota)(c_p+\iota)\mathcal{D}\right]\\
&=\left[ (1+\alpha+\overline{a}_\mathrm{He})c_p+(c_v+\iota)(c_p+\iota)
\frac{(1-\alpha)\alpha}{2-\alpha}\right], \nn \\
\left(\frac{\partial w}{\partial \rho}\right)_{\!\!T}
&=-\frac{T}{M^*}(c_p+\iota)\frac{\mathcal{D}}{M^*n_\rho} \\
&=-\frac{T}{M^*}\frac{1}{M^*n_\rho}
\left[ (c_p+\iota)
\frac{(1-\alpha)\alpha}{2-\alpha}\right].\nn
\end{align}

Then the Jacobian~\cite[Eq.~(9)]{ApJ:24}
\be
 \mathcal{J} \equiv 
\frac{\partial (w,p)}{\partial (T,\rho)} 
= 
\left(\frac{\partial w}{\partial T}\right)_{\!\! \rho}
\left(\frac{\partial p}{\partial \rho}\right)_{\!\! T}
-\left(\frac{\partial w}{\partial \rho}\right)_{\!\! T}
\left(\frac{\partial p}{\partial T}\right)_{\!\! \rho}
\label{Jacobian_up} 
\ee
after some algebra reads
\begin{widetext}
\be
\mathcal{J} =\frac{T\mathcal{N}_\mathrm{tot}}{(M^*)^2}
\left\{c_p\mathcal{N}_\mathrm{tot}+
\left[(c_v+\iota)(c_p+\iota)+\iota\right]
\mathcal{D}
\right\}
=\frac{(1 \! + \! \alpha \! + \! \overline{a}_\mathrm{He})T}{(M^*)^2}
\left[ c_p(1 \! +\! \alpha \! + \! \overline{a}_\mathrm{He})
+
(c_vc_p+2c_p \iota+\iota^2)
\frac{(1-\alpha)\alpha}{2-\alpha}
\right].
\label{Jacobian_analytical}
\ee
\end{widetext}
In such a way we obtained that for the 2-component cocktail
\be
\mathcal{J}=\frac{T}{M^*}
(\mathcal{N}_\mathrm{tot}-\mathcal{D})\mathcal{C}_p.
\label{simple_J}
\ee
Representing the sound velocity at evanescent frequency when
ionization degree follows it equilibrium values determined by Saha equation
we have
\begin{align}
c_0^2\equiv\left(\frac{\partial p}{\partial \rho}\right)_s=\gamma_0\frac{p}{\rho},
\end{align}
The general result~\cite[Eq.~(9)]{ApJ:24}
\be
\gamma_0=\frac{\rho}{p}\cdot\frac{\mathcal{J}}{\mathcal{C}_p}
\ee
after substitution $\mathcal{J}$ from \Eqref{simple_J} gives
\begin{align}
\gamma_0&=\left(1-\frac{\mathcal{D}}{\mathcal{N}_\mathrm{tot}}\right)
\frac{\mathcal{C}_p}{\mathcal{C}_v}
\label{gama_simple}
\\
&=\left[1-\frac{(1-\alpha)\alpha}{(2-\alpha)(1+\alpha+\overline{a}_\mathrm{He})}
\right]
\frac{\mathcal{C}_p}{\mathcal{C}_v}. \nn
\end{align}
This is and illustration how ionization recombination processes brock
the general relations derived for constant chemical compounds.
For solar plasma it is not strongly expressed but for interstellar plasma
it is possible that $\iota=I/T\gg1$ in this case for $\iota^2\mathcal{D}$
the heat capacity per proton to be
\begin{align}
M^*\mathcal{C}_v&\approx\iota^2\mathcal{D}\gg1,\\
M^*\mathcal{C}_p&\approx
\iota^2\frac{\mathcal{D}}{1-\mathcal{D}/\mathcal{N}_\mathrm{tot}}\gg1,
\end{align}
confer~\cite[Fig.~3 and Fig.~5]{ApJ:21,PhysA}.
In the opposite case of negligible influence of ionization processes
$\iota\ll1$ we have the trivial test for the programming of the heat capacities
\be
M^*\mathcal{C}_v\approx c_v\mathcal{N}_\mathrm{tot},\quad
M^*\mathcal{C}_p\approx c_p\mathcal{N}_\mathrm{tot},\quad
M^*(\mathcal{C}_p-\mathcal{C}_e)=\mathcal{N}_\mathrm{tot}.
\ee

The collisions of protons with protons 
are more intensive than collisions with neutral atoms and for the shear
viscosity we can use the result for completely ionized hydrogen plasma
\cite[Eq.~(43.9)]{LL10}
\begin{align}
&
\eta\approx0.4\frac{M^{1/2}T^{5/2}}{e^4 \Lambda},\qquad
\Lambda=\ln\frac{Tr_\mathrm{_D}}{e^2},\\
&
e^2=\frac{q_e^2}{4\pi\epsilon_0},\qquad
\frac1{r_\mathrm{_D}^2}=2\,\frac{4\pi e^2}{T}n_\rho\alpha,
\end{align}
where $q_e$ is the electron charge, and
$r_\mathrm{_D}$ is the Debye radius \cite[Eq.~(78.8)]{LL5}.
Analogously the electrons are scattered mainly by protons 
and \cite[Eq.~(43.10)]{LL10}
\be
\varkappa=\frac{T^{5/2}}{e^4m^{1/2} \Lambda}.
\ee
Both the kinetic coefficients $\eta$ and $\varkappa$ have weak density dependence
only through the Coulomb logarithm $\Lambda$.

For the bulk viscosity of the H-He cocktail we have recently derived 
\textit{ad hoc} results \cite{PoP:26}
\begin{align}
&
\zeta^\prime(\omega)=\frac{\zeta_0}{1+\omega^2\tau^2},
\qquad
\zeta_0=\frac{p\tau\mathcal{B}}{\mathcal{A}}, 
\qquad 
\tau=\frac{\mathcal T}{\mathcal A},\\
&
\mathcal{A}\equiv
(2-\overline{\alpha})c_v\mathcal{N}_\mathrm{tot}
+(c_v+\iota)^2(1-\alpha)\alpha,\\
&
\mathcal{B}=(1-\alpha)\,\alpha \, \iota^2/c_v,\\
&
\mathcal{T}
\equiv\tau_\mathrm{_H} c_v \overline\alpha \mathcal{N}_\mathrm{tot}.
\end{align}

The time constant $\tau_\mathrm{H}$ is probably the main detail of the present
theory. 
It describes the decay rate of an neutral hydrogen atom
\begin{align}
\frac1{\tau_\mathrm{_H}}=\beta n_e,
\quad \beta(T) =\langle v_e\sigma(\varepsilon_e)\rangle,
\quad\varepsilon_e=\frac12mv_e^2.
\end{align}
Here brackets denote averaging with respect of Maxwell distribution.
Including here the ionization cross section
\begin{align}
&
\sigma(\varepsilon_e)\sim (\varepsilon_e-I)^{\mathrm{w}},\\
&
(\varepsilon_e-I)\sim T\ll I, \qquad
\mathrm{w} \approx 1.18\sim 1
\end{align}
by Wannier~\cite{Wannier:53}
we use the approximation for the rate of ionization reaction 
\begin{align}
 \beta(T)
 &\approx \frac{2}{\sqrt{\pi}}C_\mathrm{W} \Gamma(\mathrm{w}+1)\frac{\e^{-\iota}}{\iota^{\mathrm{w}-1/2}}
 \beta_\mathrm{B},
 \label{beta_a}\\
\beta_\mathrm{B}&\equiv v_\mathrm{_B} a_\mathrm{_B}^2
=6.126\times 10^{-15}\, \mathrm{m^3/s},
\qquad I=\frac{e^2}{ 2a_\mathrm{_B}},
 \nn\\
  v_\mathrm{_B}&=c\,\frac{e^2}{\hbar c},\quad 
  a_\mathrm{_B}=\frac{\hbar}{mc}\cdot\frac{\hbar c}{e^2},\quad
  \frac{e^2}{\hbar c}\approx\frac1{137}.
\end{align}
In the above expressions for the Bohr velocity $ v_\mathrm{_B}$ 
and Bohr radius $a_\mathrm{_B}$,
$c$ denotes light velocity representing the Sommerfeld 
fine structure constant.
The constant
$C_\mathrm{W} \approx 2.7$ is evaluated after 
the experimental study~\cite[Fig.~6]{McGowan:68}.

For the Mandelstam-Leontovich time constant we finally obtain
\begin{align}
\tau & = \frac{1}{\beta n_\rho}\cdot
\frac{1}{2-\alpha}\cdot
\dfrac{c_v}
{c_v+(c_v+\iota)^2 \mathcal{D}/\mathcal{N}_\mathrm{tot}} \\
&=\frac1{\beta n_\rho} \,
\frac{c_v}
{c_v(2-\alpha)(1+\alpha+\overline{a}_\mathrm{He})+(c_v+\iota)^2(1-\alpha)\alpha} \nn \\
& = \frac{1}{\beta n_\rho}\frac{1}{2-\alpha} \dfrac{c_v}{\tilde{c}_v},
\quad
\tilde{c}_v \equiv \frac{M^*}{\mathcal{N}_\mathrm{tot}} \mathcal{C}_v = 
c_v+ (c_v +\iota)^2 \dfrac{\mathcal{D}}{\mathcal{N}_\mathrm{tot}} \nn
\end{align}
in agreement for $\overline{a}_\mathrm{He}=0$ with the result for pure
hydrogen plasma~\cite[Eqs.~(75) and (113)]{PhysA}, where
$M^*\mathcal{C}_v $ is the heat capacity per hydrogen atom at fixed volume;
$M^*\mathcal{C}_v/\mathcal{N}_\mathrm{tot}$ is the averaged heat capacity
per particle of the cocktail;
and $M^*\mathcal{C}_v/\mathcal{N}_\mathrm{tot}c_v$ is its relative value.
In Ref.~\cite{PhysA} Mandelstam-Leontovich time constant was derived
by application of Boltzmann $H$-theorem for hydrogen plasma for which
ionization is initially slightly different from the equilibrium value while
in the recent study Ref.~\cite{PoP:26} was analyzed the complex 
generalized compressibility. 
The agreement of these different approaches is a hint that the
used formula for $\tau$ is reliable.

For $\overline{a}_\mathrm{He}>0$, the comparison is also straightforward as $\mathcal{N}_\mathrm{tot} = 1 + \alpha + \overline{a}_\mathrm{He}$ here and therefore derived general results is the same. 
And this shows the algorithm for addition of  more noble chemical elements not participating in the ionization-recombination processes.
Let us analyze also a special case of an almost completely ionized hydrogen
$\alpha\approx 1$. 
In this case $\mathcal{D}\approx 0$, 
and for the electron density we have $n_e\approx n_\rho$
meaning that at these conditions $\tau\approx \tau_\mathrm{H}$ i.e.
\be
\frac{1}{\tau}\approx \beta n_e.
\ee
In this special case $1/\tau$ is the decay rate of the last remaining
neutral hydrogen atoms with respect of the electron impacts.

For high frequencies $\omega\tau\gg1$ the damping rate related to the second
viscosity is dispersion-less
\be
k_\infty\equiv
k^{\prime\prime}(\omega\rightarrow\infty)
=\frac{\omega^2\zeta^\prime(\omega)}{2\rho c^3}
\approx\frac{1}{2\tau c_\infty}\frac{\mathcal{B}}{\gamma_a\mathcal{A}},
\label{damping_inf} 
\ee
where for the sound velocity at these high frequencies
we have
\be
c=c_\infty=\sqrt{\frac{\gamma_a p}{\rho}},
\qquad
\gamma_a=\frac{c_p}{c_v}=\frac53.
\ee

\subsection{Solar chromospheric profiles of the main notions}

Using the height dependent profiles of the temperature $T^\prime$ and mass density $\rho$~\cite[Model C7]{Avrett:08}~(AL08), the main introduced notions are calculated and analyzed in this subsection.

\begin{figure}[ht]
\centering
\includegraphics[scale=0.5]{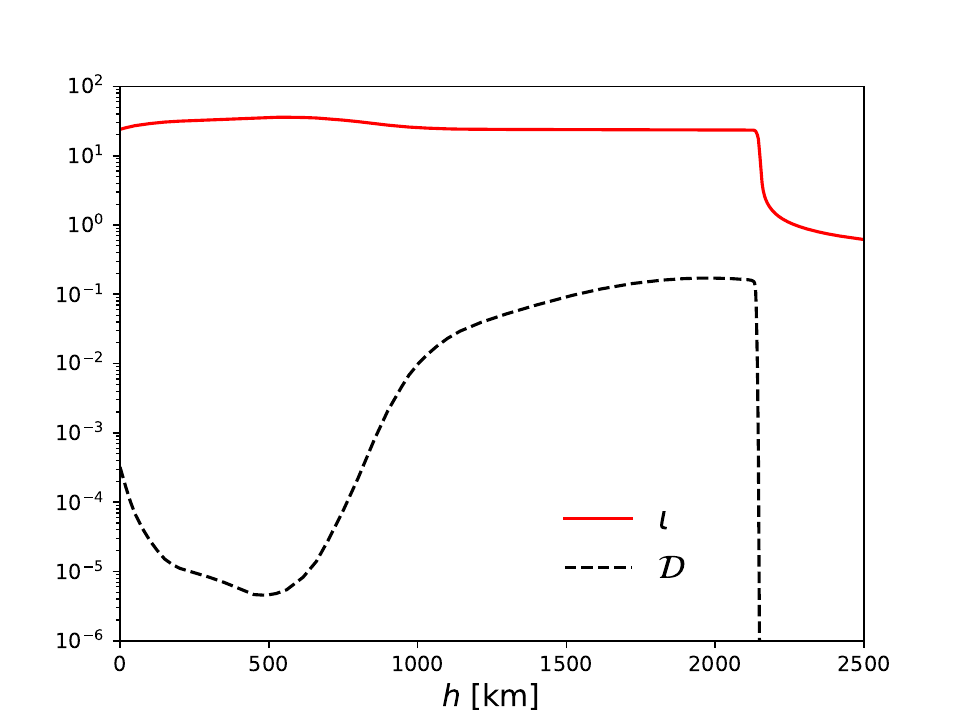}
\caption{The height dependence of dimensionless function $\lg\mathcal{D}(h)$
\Eqref{cal_D} which is via \cite[Model C7]{Avrett:08}
for heights $h \equiv x <2.1$~Mm.
This function is the main ingredient of all analytical results in which
ionization-recombination processes are relevant.
For comparison in the same logarithmic ordinate
is given the profile of the dimensionless variable $\iota(h)\gg1$.
The abrupt change of both variables is physically in the solar transition region and it is beyond the scope of the present study.
}  
\label{fig:D}
\end{figure}

First, the height profile of the function $\mathcal{D}(h)$
is depicted in \Fref{fig:D} below the profile of $\iota(h)$. 
Both these notions are central ingredients in our analytical results.
As $\mathcal{D} \rightarrow 0$ in the solar transition region, $\zeta \rightarrow 0$ and $\iota < 1$ in the coronal conditions beyond the current study.

\begin{figure}[ht]
\centering
\includegraphics[scale=0.5]{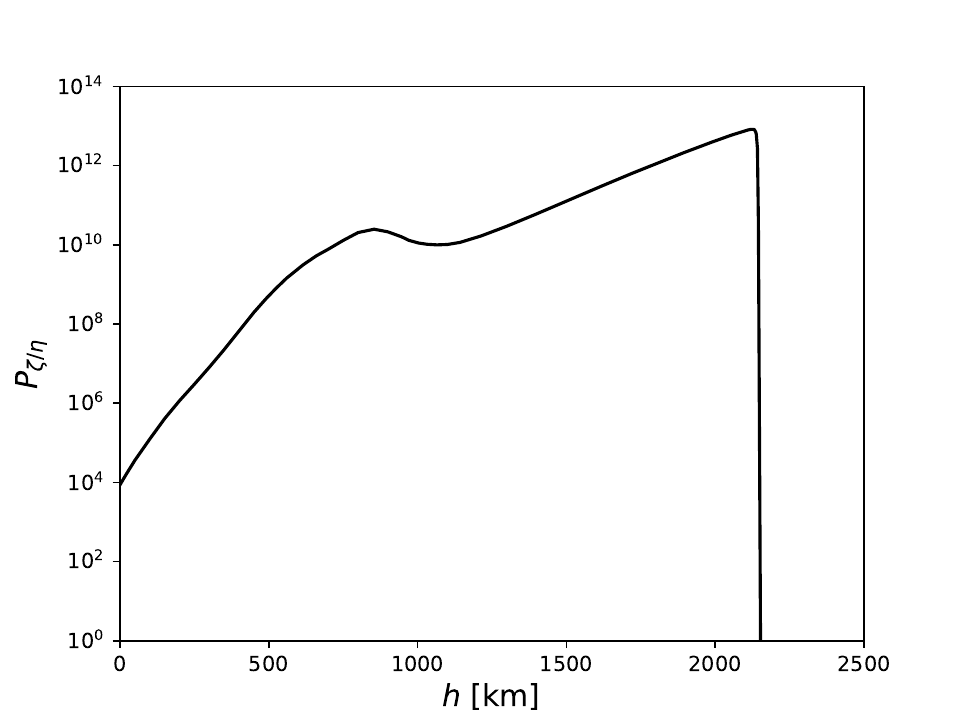}
\caption{The height $h$ profile of bulk viscosity Prandtl number
\Eqref{visc_Prandtl_}
$\mathrm{P}_{\zeta/\eta}\equiv\zeta/\eta$ via \cite[Model C7]{Avrett:08}.
One of the purposes of the present study is to emphasize that
for heights $h \equiv x <2.1$~Mm from the solar photospere the bulk viscosity indispensable must be included in the considerations of acoustic wave heating of chromosphere.
}  
\label{fig:prandtl}
\end{figure}

The comparison between different sound waves damping mechanisms
represented in \cite[Fig.~1]{PoP:20} for the solar atmosphere via AL08 reveals
that the bulk viscosity is many orders of magnitude larger than the shear one in \Fref{fig:prandtl}.
\be
\mathrm{P}_{\zeta/\eta}\equiv\frac{\zeta_0}{\eta}\gg 1.
\label{visc_Prandtl_}
\ee
The height dependence of bulk viscosity Prandtl number
is drawn in \Fref{fig:prandtl}.
For high enough frequencies the viscosity terms are equalized 
$\zeta_0/(\omega_c\tau)^2=\eta$  and above $\omega_c$ 
the frequency independent shear viscosity is larger.
For our example
for $\omega_c \approx 35\times 10^{-3}\,\mathrm{s}^{-1}$ 
and $f_c=\omega_c/(2\pi)\approx 5.5\,\mathrm{mHz}.$
\begin{figure}[ht]
\centering
\includegraphics[scale=0.5]{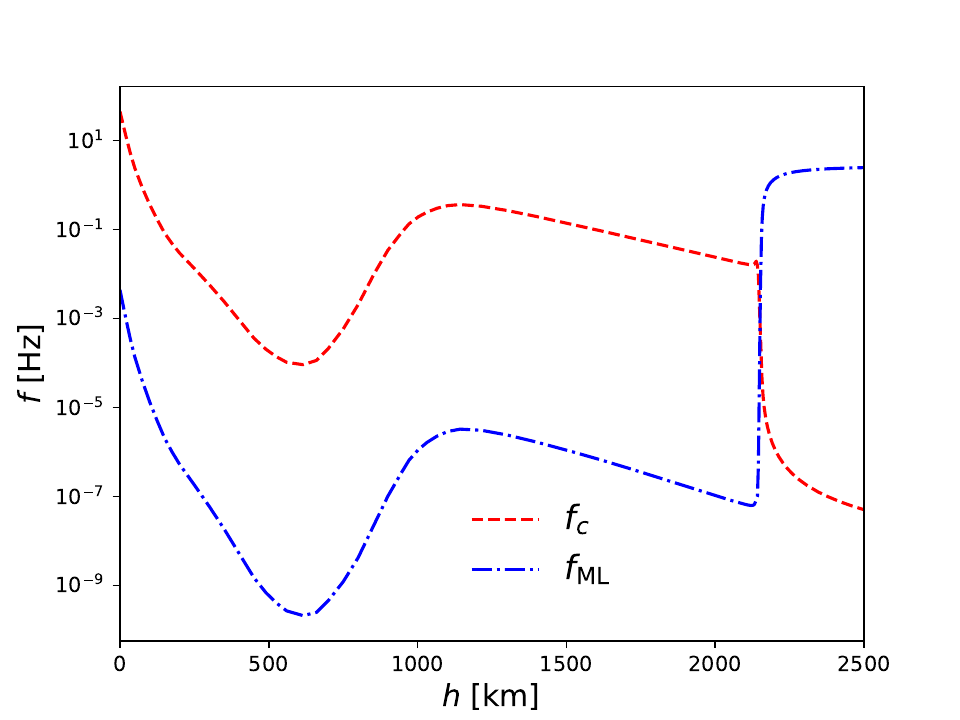}
\caption{Height $h$ profiles of the frequencies 
$f_\mathrm{ML}\equiv 1/2\pi\tau
\ll f \ll 
f_c\equiv f_\mathrm{ML}\sqrt{\mathrm{P}_{\zeta/\eta}}$ again via \cite[Model C7]{Avrett:08}.
The damping rate $k_\infty$ is frequency independent 
according to \Eqref{damping_inf}  and determined by bulk viscosity $\zeta$
and volume heating.
In this frequency interval the acoustic wave heating according to
\Eqref{force_density_and-heating rate}
is proportional to the total energy flux of sound waves.
}  
\label{fig:freqs}
\end{figure}
After so introduced motions we can address to the solution of
\Eqref{system}.

\subsection{Equations for temperature and density profiles}

The system of equations \Eqref{system} can be rewritten
as
\begin{align}
\begin{pmatrix}
\md_x w\\
\md_x p
\end{pmatrix}
=\begin{pmatrix}
\left(\dfrac{\partial w}{\partial T}\right)_{\!\! \rho}
&\left(\dfrac{\partial w}{\partial \rho}\right)_{\!\! T}\\
\left(\dfrac{\partial p}{\partial T}\right)_{\!\! \rho}&
\left(\dfrac{\partial p}{\partial \rho}\right)_{\!\! T}
\end{pmatrix}
\begin{pmatrix}
\md_xT\\
\md_x\rho
\end{pmatrix}
=\begin{pmatrix}
g\\
f
\end{pmatrix}
.
\end{align}
And we can express the derivatives of the profile
\begin{align}
\begin{pmatrix}
\md_xT\\
\md_x\rho
\end{pmatrix}
=
\frac1{\mathcal{J}}
\begin{pmatrix}
\quad \left(\dfrac{\partial p}{\partial \rho}\right)_{\!\! T} &
-\left(\dfrac{\partial w}{\partial \rho}\right)_{\!\! T}\\
-\left(\dfrac{\partial p}{\partial T}\right)_{\!\! \rho} &
\quad \left(\dfrac{\partial w}{\partial T}\right)_{\!\! \rho}
\end{pmatrix}
\begin{pmatrix}
g\\
f
\end{pmatrix}.
\label{final_system}
\end{align}

This system has obvious solution
\begin{align}
&
\left.
\begin{pmatrix}
T\\
\rho
\end{pmatrix}
\right\vert_{x_0}^{x_f}
=\int\limits_{x_0}^{x_f}
\begin{pmatrix}
\quad \left(\dfrac{\partial p}{\partial \rho}\right)_{\!\! T} g&
-\left(\dfrac{\partial w}{\partial \rho}\right)_{\!\! T}f\\
-\left(\dfrac{\partial p}{\partial T}\right)_{\!\! \rho} g&
\quad \left(\dfrac{\partial w}{\partial T}\right)_{\!\! \rho}f
\end{pmatrix}
\frac{\md x}{\mathcal{J}(x)}.
\label{profile_system}
\\
&
\left.\ln q(x)\right\vert_{x_0}^{x_f}=-\int_{x_0}^{x_f} (2k_\infty^{\prime\prime})\,\md x.
\end{align}
For the initial integration point can be chosen the surface of photosphere $x_0=0$
and for final point $x_f$ 
can be chosen some height slightly below the transition region (TR) where
the second viscosity is small due to low density $n_\rho$ and high ionization
degree $\alpha(x_f)\approx 1$.
The solution of the system depends in two unknown parameters
The mass debit $j$ and the acoustic energy flux on the surface of photosphere
$q_0=q(x=x_0)$ are two unknown parameters which can be determined by
two temperatures $T_0=T(x_0)$, $T_f=T(x_f)$ and densities
$\rho_0=\rho(x_0)$, $\rho_f=\rho(x_f)$.
For this fitting procedure we can use in the integration in
\Eqref{profile_system} the observed profiles $T(x)$ and 
$\rho(x)=(n_p+n_0)M^*$ taken from Avrett and Loeser~AL08.
Then having no freedom we can taste whether the profiles are consistent
if using so determined $j$ and $q_0$ we can solve \Eqref{profile_system} 
as a system of ordinary differential equations.
If we have qualitative agreement, any omitted in the first approximation terms
can be included as perturbation in the right side of the equations, for example, 
the influence of heat conductivity and bulk viscosity on the solar wind
\begin{align}
q&\rightarrow q-\md_x(\varkappa\,\md_xT)+\zeta (\md_x U)^2,\\
f&\rightarrow f+\md_x(\zeta\md_xU).
\label{perturbation_by_solar wind}
\end{align}

\section{Radiative cooling}

Radiative cooling is an important ingredient in the equation for the 
temperature $T(x)$ and density profiles $\rho(x)$  
\Eqref{system} and \Eqref{profile_system}.

\begin{figure}[ht]
\centering
\includegraphics[scale=0.5]{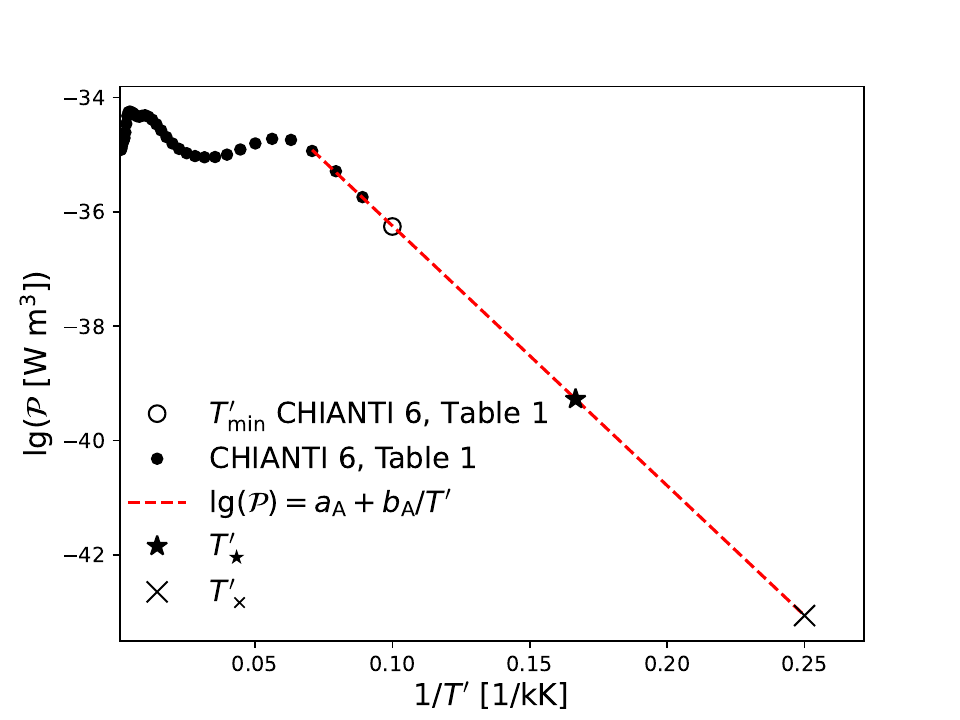}
\caption{Radiative loss rate $\lg \mathcal{P}$
versus $1/T^\prime$; point are after~\cite[Table~1 and Fig.~5]{Dere:09}, photospheric abundance.
The 4 smallest temperature points excellently approximate a straight line.
For comparison are marked the lowest temperature in the chromosphere
$T_\times^\prime= 4$~K and the photosphere temperature 
$T_\star^\prime= 6$~kK.
}  
\label{fig:arrhenius-fit}
\end{figure}

\begin{figure}[ht]
\centering
\includegraphics[scale=0.5]{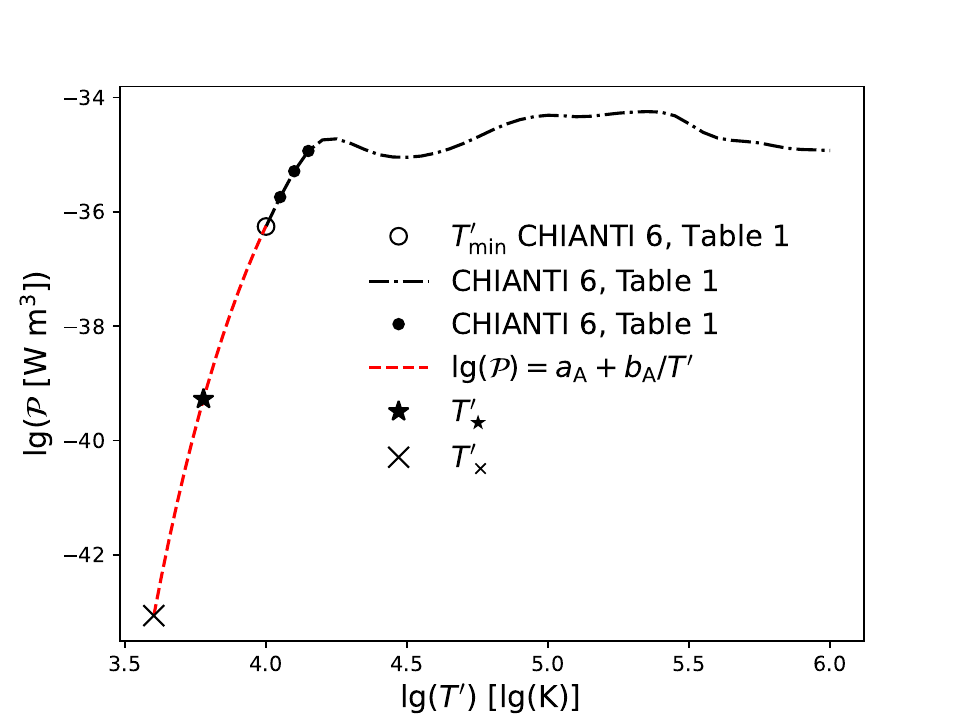}
\caption{Radiative loss rate $\lg \mathcal{P}$ versus $\lg T^\prime$ with CHIANTI~6 \cite{Dere:09}
The dashed line is the Arrhenius extrapolation 
\Eqref{Arrhenius_extrapolation} which used for $T^\prime<T_\mathrm{min}^\prime=10$~kK, while the dash-dotted line is the CHIANTI~6 radiative loss rate.
}  
\label{fig:rad-extrapolation}
\end{figure}

The radiation cooling power per unit volume factorizes to 
a product of a temperature dependent
function and product of electron and proton concentrations
\be
\mathcal{Q}_r=\mathcal{P}(T)n_en_p.
\ee
For the low temperature H-He cocktail
$n_p\approx n_e \approx n_\rho\alpha$.
Available data for the energy loss function can be found in 
Ref.~\cite[Table~1 and Fig.~5]{Dere:09} (CHIANTI~6) for photospheric abundance, however this function is tabulated from $T^\prime_\mathrm{min} = 10$~kK reproduced in \Fref{fig:arrhenius-fit} but with reciprocal temperature dependence $1/T^\prime$.
The chromospheric temperatures are even smaller than the energy difference
of the hydrogen atom
\be
\hbar\omega_{23}=\frac{\mathrm{R}}{2^2}-\frac{\mathrm{R}}{3^2}
\approx 1.9\,\rm{eV}\approx 22\,kK
\ee
and in this case the tail of the Maxwell distribution of the electrons can dominantly activate only the lowest radiation level 
\be
E_2=\frac{\mathrm{R}}{2^2} 
\ee
with activation energy
\be
\hbar\omega_{12}=\frac{\mathrm{R}}{1^2}-\frac{\mathrm{R}}{2^2}
\approx 10.2\,\rm{eV}.
\ee
This qualitative consideration can be easily seen in the Arrhenius plot
\Fref{fig:arrhenius-fit}
where the lowest 4 temperature points lie on a straight line.
The linear regression of these 4 lowest temperature points gives
the Arrhenius extrapolation to lower temperatures
\be
\mathcal{P}_\mathrm{A}(T)=\mathcal{P}_0\,\e^{-E_\mathrm{A}/T}
\label{Arrhenius_extrapolation}
\ee
to low temperatures 
$T<T_\mathrm{\min}^\prime= 10$~kK.
The fitted line slope of gives the Arrhenius (or the activation) energy
\be
E_\mathrm{A}
=\kb\times 104.6~\mathrm{kK}=9.1\,\mathrm{eV}\sim \hbar\omega_{12}.
\ee
Qualitatively this interpretation is given in Ref~\cite[Figs~9 and 10]{Landini:90} where it is pointed out that the low temperature maximum of the function $\mathcal{P}(T)$ is mainly determined by emission of the Lyman series.
The pre-exponential factor of the Arrhenius approximation
\be
\mathcal{P}_0 \approx 2 \times 10^{-32}~\mathrm{W\,m^3},
\;\; \rm{1\,erg\, cm^3/s=10^{-13}\,W\,m^3}
\ee
and using \Eqref{Arrhenius_extrapolation} with the calculated from the Arrhenius fit parameters, the extrapolation to temperatures smaller than $T^\prime_\mathrm{min}$ is shown in \Fref{fig:rad-extrapolation} alongside the CHIANTI~6 tabulated energy loss function.
Finally, it should be noted that the $1/T^\prime$ linear fit is performed on the lg scale and consequently rescaled to the ln scale obviously due to the Arrhenius dependence~\Eqref{Arrhenius_extrapolation}.
Perhaps the simplest evaluation of the low temperature behavior of the energy loss
function is to accept that for low temperatures $T\ll\hbar\omega_{23}$ we have
$E_\mathrm{A}=\hbar\omega_{12}$ and the activation exponent 
$\e^{-E_\mathrm{A}/T}=\e^{-0.75\,\iota}$ and in \Eqref{Arrhenius_extrapolation}
to use 
$\mathcal{P}_0=\exp(0.75\,\mathrm{R}/T_\mathrm{min})$.

Having all ingredients we can describe numerical solution of the temperature 
profiles described in the nest section.

\section{Numerical solution}

\begin{figure}[ht]
\centering
\includegraphics[scale=0.5]{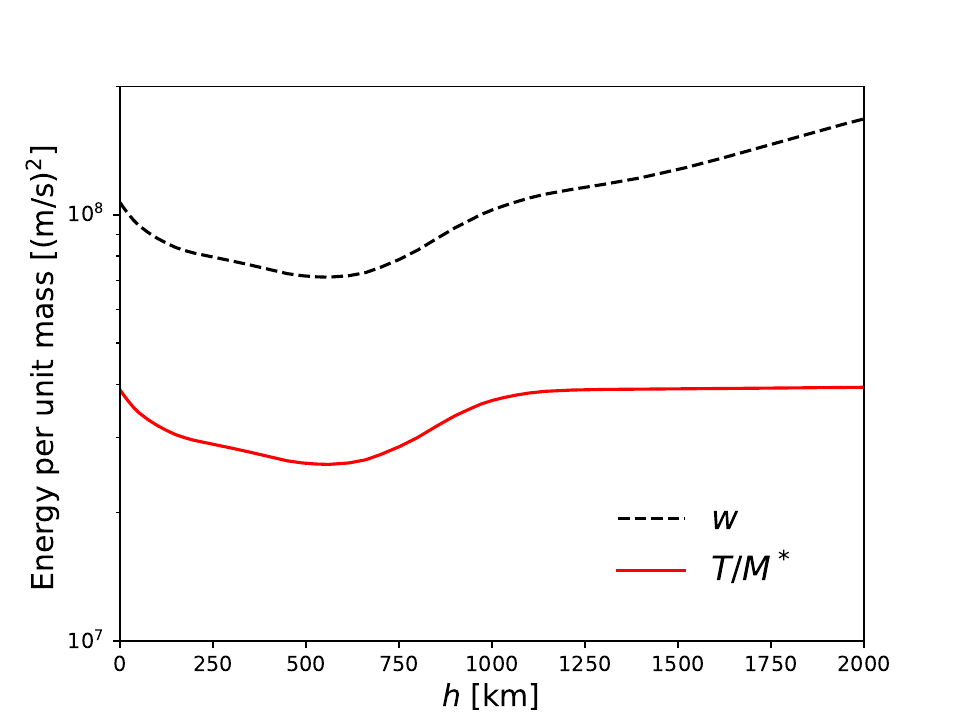}
\caption{Height dependency (dashed line) of the enthalpy 
per unit mass $w$ having dimension m$^2$/s$^2$ via \cite[Model C7]{Avrett:08}.
The variable $T/M^*$ lower (solid line) is almost the square of the
thermal velocity of protons $v_{Tp}=\sqrt{T/M}$; see the notations
in \cite{LL10}.
Both variables have a broad minimum at 
$h_\mathrm{min}\approx 560$~km.
We use the slope at $h=0$ and the position of the minimum in order to determine
the fluxes of mass $j$ and acoustic waves $q_0$.
The solutions of the general system \Eqref{final_system} reproduces
these properties.
}  
\label{fig:enthalpy}
\end{figure}

The first step of the launching of the new theory is to determine the indispensable
parameters of the theory illustrating its main ingredients the fluxes of the mass $j$
and acoustic energy $q_0$ at $x=0$.
The height profile of the enthalpy $\tilde w=\frac12U^2+w+gx$ 
per unit mass is depicted in \Fref{fig:enthalpy} via AL08;
for the parameters of solar atmosphere kinetic $\frac12U^2$ and 
potential $gx$  energy are negligible.
Starting our analysis at the photospheric surface $x=0$,
for dense cold plasma the kinetic coefficients in the formula for wave damping
\Eqref{complete_damping} are negligible $k^{\prime\prime}(x=0)\approx 0$ and
the decreasing of the enthalpy at photosphere surface according \Eqref{enthalpy_derivatives} determines the wind velocity $U_0=U(x=0)$ and the debit $j=U_0\rho_0$ at photospheric surface
\begin{align}
&
U_0=-\dfrac{\mathcal{Q}_r(x=0)}{\rho_0\left.\dfrac{\md w}{\md x}\right\vert_{x=0}}\approx 17\mathrm{\, \frac{cm}{s}},\\
&
j=\rho_0U_0 =50\mathrm{\,\frac{mg}{m^2\,s}}.
\end{align}
The small value of the wind velocity $U(x)$ demonstrates
that convective acceleration $a_\mathrm{conv}=U\md U$ is small 
in lower photosphere justifies we use hydrostatic approximation in
\Eqref{system} momentum equation.

Slightly above the surface at height 
$x_\mathrm{min}\approx 560$~km enthalpy $w$ has a minimum
and the radiation cooling in \Eqref{enthalpy_derivatives} is compensated 
by bulk viscosity heating
$\mathcal{Q}_r(x_\mathrm{min})=\mathcal{Q_\zeta}(x_\mathrm{min})$
where from \Eqref{force_density_and-heating rate} we evaluate
\be
q_0 \equiv q(x=0)\approx\frac{\mathcal{Q}_r(x_\mathrm{min})}
{2k_\infty^{\prime\prime}(x_\mathrm{min})}
\approx 320\,\mathrm{\frac{kW}{m^2}}.
\label{wave_energy_flux}
\ee
The AL08 height profiles for temperature $T(x)$ and density $\rho(x)=M^*(n_0(x)+n_p(x))$ are used for these calculations,
the extrapolation formula for the radiation loss function 
$\mathcal{P}_\mathrm{A}(T(x))$
\Eqref{Arrhenius_extrapolation} and the result for the height frequency
bulk viscosity damping rate 
\Eqref{damping_inf}.
The order evaluation of wave energy flux according of above
\Eqref{wave_energy_flux}
is in the same order with other evaluations and it is a hind that
we are approaching to the final solution of the problem.

Now we have no freedom. 
The parameters determined by the height minimum of the enthalpy must be used to describe the whole profile which is our next task.
First step is to check the consistency of our theory at small heights.
We can use the AL08 profiles to substitute them in the right
side of \Eqref{profile_system} in order to check whether they can describe
the minimum of the temperature $T(x)$ close to $x_\mathrm{min}$.
In case of acceptable agreement, we can calculate absolutely new profiles
$T(x)$ and $\rho(x)$ using the system \Eqref{final_system}.
If necessary in sequential approximation
we can take into account the influence of dissipation coefficients in the solar wind
\Eqref{perturbation_by_solar wind}.
The scheme can be extended to derivation of the equations of the frequency dependent spectral density of the acoustic waves and their three dimensional motion.
The possible complications are infinite.
But the purpose of the present study is strictly limited.
The goal of our work is to open the Pandora box by
including the influence of the bulk viscosity $\zeta$ in the eternal problem
of heating of the solar chromosphere.
We stress out the huge value of the bulk viscosity Prandtl number
depicted in \Fref{fig:prandtl}.
This significant value change the conclusion of the former
researches of the acoustic heating of the chromosphere:
\textit{``The inferred wave energy fluxes based on our observations are not sufficient to maintain the solar chromosphere''}~\cite{Molnar:23} to the opposite:
\textbf{the wave energy fluxes based on our observations are sufficient to maintain the solar chromosphere}.
As the heating profiles $T(h)$ and $\rho(h)$ depend on fluxes 
of mass $j$ and acoustic energy $q_0$ coming from the photosphere,
we arrive at the conclusions that heating depends on the boundary conditions,
confer~\cite{Judge:24} where the opposite statement is concluded:
\textit{``Heating depends on the state of the corona, not simply on boundary conditions''}~\cite{Judge:24}.

The results in this study were obtained using Fortran program and modules and the figures were prepared with Python and its Matlplotlib library~\cite{Hunter:07}.


\section{Conclusions and perspectives}
The problem of heating of the solar chromosphere is 
in the agenda of the astrophysics for at least half a century.
The hydrodynamic and in general the MHD approach are used in uncountable numerical simulations of the solar atmosphere.
It is strange that the huge value of the bulk viscosity
$\mathrm{P}_{\zeta/\eta}\sim 10^{10}$ has not been taken into account up to now;
neglecting a giraffe on the background of an atom.
The purpose of the present article is to focus the attention on the possible
last forgotten detail in the problem of chromosphere heating.
Starting with \textit{ab initio} calculated ionization cross-section by Wannier in 1953~\cite{Wannier:53}, we have calculated the bulk viscosity of a realistic for solar atmospheric H-He cocktail
and is such a way have opened the perspective \textit{ab initio} to describe 
the problem of the chromosphere heating. 

As the volume viscosity $\zeta$ for partially ionized plasma creates the most intensive damping of acoustic waves, it has already become an indispensable ingredient in every consideration of the problem of the chromosphere heating.
For the initial illustration, we represent only one dimensional, short wavelength,
static approximation which explains two important details of the temperature profile:
initial decreasing of the temperature slightly above the chromospheric surface and
the temperature minimum when weak ionization $0 \lesssim \alpha\ll1$ starts creating
the volume viscosity $\zeta$. 
No doubts further numerical calculations can explain the significant heating in various solar and stellar phenomena.
But the theoretical physics in the beginning must have only qualitative
correspondence to the experiment and observations.
Hopefully, later on all details will be incorporated in a coherent picture.

\acknowledgments
The authors are thankful to Diana~Bakkar, Yoana~Ruseva, Nikolay~Aleksandrov, Emil~Petkov, Stefan~Stefanov, Valya Mishonova and Morena~Angelova, 
for their interest in this study,
to Iglika~Dimitrova for the collaboration in the early stages of the research~\cite{PhysA,ApJ:21}.

\bibliography{zeta}

\end{document}